\documentclass[usenatbib]{mn2e}
\usepackage{ulem}
\usepackage{amsmath}
\usepackage{graphicx}
\usepackage{subfigure}
\usepackage{color}
\begin{document}
\title[Cosmic Rays in Arp 220]{Cosmic Rays, $\gamma$-Rays, and Neutrinos in the Starburst Nuclei of Arp 220}

\author[Yoast-Hull, Gallagher, \& Zweibel]{Tova M. Yoast-Hull$^{1,2}$\thanks{E-mail: yoasthull@wisc.edu}, John. S. Gallagher III$^3$, and Ellen G. Zweibel$^{1,2,3}$\\
$^1$Department of Physics, University of Wisconsin-Madison, WI, USA, 53706\\
$^2$Center for Magnetic Self-Organization in Laboratory and Astrophysical Plasmas, University of Wisconsin-Madison, WI, USA, 53706\\
$^3$Department of Astronomy, University of Wisconsin-Madison, WI, USA, 53706}

\maketitle

\begin{abstract}
The cores of Arp 220, the closest ultraluminous infrared starburst galaxy, provide an opportunity to study interactions of cosmic rays under extreme conditions.  In this paper, we model the populations of cosmic rays produced by supernovae in the central molecular zones of both starburst nuclei.  We find that $\sim65 - 100\%$ of cosmic rays are absorbed in these regions due to their huge molecular gas contents, and thus, the nuclei of Arp 220 nearly complete proton calorimeters.  As the cosmic ray protons collide with the interstellar medium, they produce secondary electrons that are also contained within the system and radiate synchrotron emission.  Using results from $\chi^2$ tests between the model and the observed radio spectral energy distribution, we predict the emergent $\gamma$-ray and high-energy neutrino spectra and find the magnetic field to be at milligauss levels. Because of the extremely intense far-infrared radiation fields, the $\gamma$-ray spectrum steepens significantly at TeV energies due to $\gamma$--$\gamma$ absorption.
\end{abstract}


\section{Introduction}

Arp~220 is the nearest ($d =$ 77~Mpc) example of an ultraluminous infrared galaxy (ULIRG) that supports star formation at extreme levels. It contains two nuclei separated by 350~pc, both surrounded by massive discs of dense molecular gas \citep[e.g.,][]{Smith98,Downes98,Mundell01,Sakamoto08,Scoville15}. Radio detections of supernovae at a rate of  1--3 yr$^{-1}$ \citep{Smith98,Rovilos05,Lonsdale06,Parra07} confirm that huge populations of massive stars are present with an implied star formation rate (SFR) of $>100~M_{\odot}$~yr$^{-1}$. Although Arp 220 could contain active galactic nuclei (AGNs), particularly in the western nucleus, the observed supernova rates indicate that star formation provides a substantial fraction of the power radiated by the nuclei.

The nuclei of Arp~220 provide access to the high-intensity mode of star formation in dense molecular media that appears to have been more common in young galaxies. These types of environments are of special interest from a range of perspectives, including the information they can provide regarding the role of galactic winds, cosmic rays, and magnetic fields in feedback processes that influence galaxy evolution. Previous investigations show that Arp~220 is likely to be a hadronic cosmic ray calorimeter where all of the power in cosmic rays is absorbed within the nuclear starburst zones \citep[e.g.,][]{Torres04}. Both nuclei also contain extremely intense far-infrared (FIR) radiation fields \citep[e.g.,][]{Soifer00,Barcos15,Scoville15}, and the west nucleus is optically thick in the FIR to wavelengths of $\lambda \approx 800~\mu$ \citep{Sakamoto08,Papadopoulos10}. The production of the observed radio synchrotron emission then requires magnetic fields of milligauss strength \citep[e.g.,][]{Lisenfeld96,Lacki10b}. 

%
%
\begin{table*}
\begin{minipage}{170mm}
\begin{center}
\caption{Input Model Parameters}
\begin{tabular}{llllc}
\hline
Physical Parameters & East Nucleus & West ST & West CND & References\\
\hline
Distance & 77.0 Mpc & 77.0 Mpc & 77.0 Mpc & \\
CMZ Radius & 70 pc & 90 pc & 30 pc & 1,2,3\\
CMZ Disc Scale Height$^{a}$ & 40 pc & 40 pc & 40 pc & 4\\
Molecular Gas Mass & $6 \times 10^{8}$ $M_{\odot}$ & $4 \times 10^{8}$ $M_{\odot}$ & $6 \times 10^{8}$ $M_{\odot}$ & 2,5 \\
Ionized Gas Mass$^{b}$ & $3 \times 10^{6}$ $M_{\odot}$ & $2 \times 10^{6}$ $M_{\odot}$ & $3 \times 10^{6}$ $M_{\odot}$ & \\
Average ISM Density$^{c}$ & $\sim$7700 cm$^{-3}$ & $\sim$3500 cm$^{-3}$ & $\sim$42 000 cm$^{-3}$ & \\
FIR Luminosity & $3\times 10^{11}$ $L_{\odot}$ & $3\times 10^{11}$ $L_{\odot}$ & $6\times 10^{11}$ $L_{\odot}$ & 2\\
FIR Radiation Field Energy Density$^{d}$ & 40 000 eV~cm$^{-3}$ & 27 000 eV~cm$^{-3}$ & 440 000 eV~cm$^{-3}$ & \\
Dust Temperature & 90 K & 50 K & 170 K & 2,6\\
SN Explosion Rate ($\nu_{\text{SN}}$) & 0.7 yr$^{-1}$ & 0.7 yr$^{-1}$ & 1.3 yr$^{-1}$ & 7\\
Star Formation Rate (SFR)$^{d}$ & 65 $M_{\odot}$ yr$^{-1}$ & 65 $M_{\odot}$ yr$^{-1}$ & 120 $M_{\odot}$ yr$^{-1}$ & \\
SN Explosion Energy$^{e}$ & 10$^{51}$ erg & 10$^{51}$ erg & 10$^{51}$ erg & \\
SN Energy in Cosmic Ray Protons$^{e}$ & 5 -- 20\% & 5 -- 20\% & 5 -- 20\% & \\
Ratio of Primary Protons to Electrons ($N_{p}$/$N_{e}$) & 50 & 50 & 50 & \\
Slope of Primary Cosmic Ray Source Function & 2.1 -- 2.3 & 2.1 -- 2.3 & 2.1 -- 2.3 & \\
\hline
\multicolumn{5}{l}{References -- (1)~\cite{Downes07}; (2)~\cite{Sakamoto08}; (3)~\cite{Aalto09}; (4)~\cite{Scoville15};}\\ 
\multicolumn{5}{l}{(5)~\cite{Downes98}; (6)~\cite{Wilson14}; (7)~\cite{Lonsdale06}.}\\
\multicolumn{5}{l}{$^{a}$Conservative estimate; \cite{Scoville15} find thinner, denser discs.}\\
\multicolumn{5}{l}{$^{b}$Assumes the ionized gas mass is roughly 0.5\% of the molecular gas mass to keep the volume filling fraction reasonably}\\ 
\multicolumn{5}{l}{less than 100\%.}\\
\multicolumn{5}{l}{$^{c}$The average ISM number density is the sum of the nuclei number densities of molecular and ionized gas, such that}\\
\multicolumn{5}{l}{$n_{\text{ISM}} = M_{\text{mol}} / (2 \times m_{p} \times \mu_{\text{mol}} \times V) + M_{\text{ion}} / (m_{p} \times \mu_{\text{ion}} \times V)$, where $\mu_{\text{mol}}$ and $\mu_{\text{ion}}$ are the mean molecular weights.}\\
\multicolumn{5}{l}{$^{d}$Derived from above parameters.}\\
\multicolumn{5}{l}{$^{e}$Excludes neutrino energy.}\\
\end{tabular}
\end{center}
\end{minipage}
\end{table*}

In this paper, we study cosmic ray interactions in the Arp~220 starburst nuclear regions using an updated version of the \citet{YoastHull13} models, hereafter YEGZ.  We develop a model with two spatial zones to accurately represent the inner and outer regions of the western nucleus as defined by its molecular gas properties \citep{Aalto09}.  We incorporate photopion energy losses and photon--photon interactions to account for the extreme FIR radiation field.  We calculate the hadronic calorimetry fraction for each nucleus for the best-fitting radio models, and we predict the total $\gamma$-ray and neutrino fluxes.

In Section 2, we review the physical parameters which we selected for the models.  Section 3 details the basic assumptions of the models and our findings for the Arp~220 starburst nuclei.  We present concluding remarks in Section 4.

\section{Arp 220 Physical Properties}

Due to its extreme properties, Arp 220 has been extensively studied across the electromagnetic spectrum.  The nuclei of Arp 220 are of particular interest as they contain more than half of the total bolometric infrared luminosity of the galaxy \citep[$1-2 \times 10^{12} ~ L_{\odot}$; e.g.,][and references therein]{Downes07,Sakamoto08}.  As the nuclei are less than 100~pc in radius, the presence of an AGN or a `hot' starburst is required to explain the extraordinarily large surface brightness in the western nucleus \citep{Downes07,Sakamoto08,Wilson14}; however, the existence of an AGN has yet to be definitively established \citep[e.g.,][]{Tunnard15}.  Further, the submillimetre observations suggest that whether or not AGNs are present, they are not the main heating source of the dust \citep[e.g.,][]{Sakamoto08}.

Estimates of the FIR luminosities of the eastern and western nuclei range from $\sim 10^{10}$ to $\sim2-3 \times 10^{11}$ and from $\sim2-3 \times 10^{11} ~ L_{\odot}$ to $\sim 10^{12} ~ L_{\odot}$, respectively \citep[e.g.,][]{Sakamoto08,Wilson14}.  The range on these luminosities is quite large due to uncertainty in the true sizes, inclinations, and opacities of the nuclei and their associated molecular disc.  To keep our adopted FIR luminosity in rough agreement with the observed supernova rate, we assume values of $3 \times 10^{11} ~ L_{\odot}$ and $9 \times 10^{11} ~ L_{\odot}$ for the eastern and western nuclei (see Table 1).  Assuming similar ratios between the nuclei for the supernova rate and molecular gas content, we adopt values of $\nu_{\text{SN}} = 0.7$ yr$^{-1}$, $M_{\text{mol}} = 6 \times 10^{8} ~ M_{\odot}$ for the eastern nucleus and $\nu_{\text{SN}} = 2.0$ yr$^{-1}$, $M_{\text{mol}} = 10^{9} ~ M_{\odot}$ for the western nucleus.  While our assumed molecular gas masses favour conservative estimates, other estimates of the gas content suggest the masses are as high as $\sim2-4 \times 10^{9} ~ M_{\odot}$ \citep{Scoville15}.

\begin{figure*}
 \subfigure[Cosmic Ray Lifetimes for Arp 220 East]{
  \includegraphics[width=0.49\linewidth]{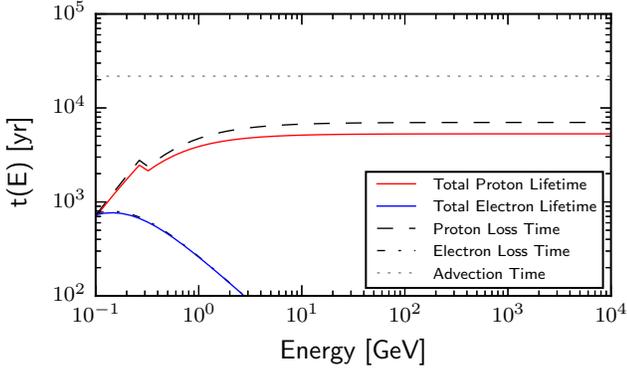}}
 \subfigure[Cosmic Ray Lifetimes for Arp 220 West CND]{
  \includegraphics[width=0.49\linewidth]{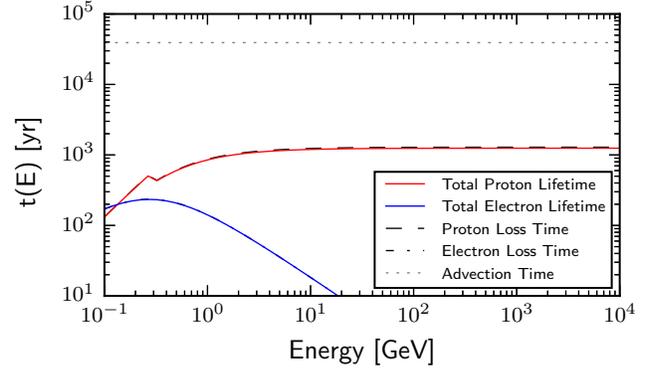}}
\caption{The total cosmic ray lifetime (solid line) is a combination of the energy loss time (dashed line) and the energy-independent advective time-scale (dotted line).  Energy losses for the cosmic ray protons (red line) include ionization, the Coulomb effect, and pion production, and energy losses for the cosmic ray electrons (blue line) include ionization, bremsstrahlung, inverse Compton, and synchrotron.  The calorimetric nature of the nuclei means that the total cosmic ray lifetime is essentially the energy loss time as the advection time is negligible.}
\end{figure*}
%
%

%
\begin{table*}
\begin{minipage}{170mm}
\begin{center}
\caption{Best-Fitting Parameters}
\begin{tabular}{lccccccc}
\hline
 & \multicolumn{3}{c}{East Nucleus} & \multicolumn{2}{c}{West ST}  & \multicolumn{2}{c}{West CND}\\
\hline
Efficiency, $\eta$ & 5\% & 10\% & 20\% & 5\% & 10\% & 5\% & 10\%\\
Spectral Index, $p$ & 2.1 -- 2.3 & 2.2 -- 2.3 & 2.3 & 2.2 -- 2.3 & 2.3 & 2.2 -- 2.3 & 2.3\\
Magnetic Field Strength, $B$ [mG] & 4.0 -- 7.5 & 2.0 -- 2.5 & 1.5 & 1.5 -- 2.5 & 1.0 & 5.25 -- 8.75 & 3.5\\
Wind Speed, $v_{\text{adv}}$ [km~s$^{-1}$] & 0--2000 & 500 -- 1400 & 1600 -- 2000 & 0 -- 2000 & 0 -- 300 & 0 -- 2000 & 0 -- 300\\
Absorption Fraction, $f_{\text{abs}}$ & 0.5 -- 1.0 & 0.6 -- 1.0 & 0.7 -- 1.0 & 0.1 -- 0.7 & 0.2 -- 0.7 & 0.1 -- 0.7 & 0.2 -- 0.7\\
Ionized Gas Density, $n_{\text{ion}}$ [cm$^{-3}$] & 400 -- 800 & 400 -- 600 & 400 -- 500 & 200 & 200 & 2000 & 2000\\
Reduced Chi-Squared, $\overline{\chi}^{2}$ & 9.6 -- 11.6 & 10.6 -- 11.5 & 11.1 -- 11.5 & 9.5 -- 11.5 & 9.9 -- 11.3 & 9.5 -- 11.5 & 9.9 -- 11.3\\
\hline
%
%
%
\end{tabular}
\end{center}
\end{minipage}
\end{table*}

CO observations of the western nucleus imply a temperature gradient increasing towards the centre and indicate significant differences in the physical conditions between the two nuclei \citep{Aalto09,Tunnard15}.  \citet{Downes07} model the western nucleus as two distinct dust sources -- a cooler (50~K) ring surrounding a hotter (170~K), dense dust core.  We use this two-zone model for the western nucleus and have adjusted our single-zone model to account for the differences in temperature and density between the two regions (see Section 3).  For the eastern nucleus, we assume a single dust temperature of 90~K \citep{Sakamoto08}.

\section{Models \& Results}

\subsection{YEGZ Model}

Previously, we developed and tested a model for cosmic ray interactions in the central molecular zones (CMZs) of star-forming and starburst galaxies \citep[YEGZ;][]{YoastHull14a, YoastHull14b}.  Our single-zone model accounts for a variety of energy losses via interactions with the interstellar medium (ISM), magnetic fields, and radiation fields and for energy-independent advective escape via a galactic wind (see Fig. 1).  The resulting cosmic ray energy spectrum depends on both the total cosmic ray lifetime and a power-law injection spectrum which is directly proportional to the volume integrated supernova rate (see YEGZ for further details).

Accounting for the production of secondary cosmic rays, we use our calculations of the population of energetic particles to predict the radio, $\gamma$-ray, and neutrino spectra.  For the $\gamma$-ray spectrum, we include both leptonic (bremsstrahlung, inverse Compton) and hadronic (neutral pion decay) emission mechanisms.  For the radio spectrum, we incorporate the effects of free--free emission and absorption \citep{YoastHull14b}.  As in our previous models, we assume that the ionized gas in the nuclei acts as a foreground screen that some fraction ($f_{\text{abs}}$) of the emitted synchrotron radiation passes through.  When the covering fraction is low ($f_{\text{abs}} \sim 0.1-0.2$), the radio spectrum flattens at low frequencies \citep{YoastHull14b}, and when the covering fraction is high ($f_{\text{abs}} \sim 1.0$), the radio spectrum turns down at low frequencies (YEGZ).

As noted above, the western nucleus in Arp 220 is best modelled with two separate regions: an inner circumnuclear disc (CND) with a surrounding torus (ST).  We model the cosmic ray populations of the two regions independently, treating each region as a uniform slab. However, the effects of absorption (free--free and $\gamma$--$\gamma$) on the resulting radio and $\gamma$-ray emission must be considered more carefully.  Absorption occurs within each emission region, and in the case of the inner CND, absorption also occurs as the emitted radiation moves through the external, ST (see the appendix for further details).  

\subsection{Radio Emission}

We perform $\chi^{2}$ tests following the approach described in YEGZ \citet{YoastHull14b}.  Comparing against radio observations for each nucleus, we vary magnetic field strength ($B$), wind speed ($v_{\text{adv}}$), ionized gas density ($n_{\text{ion}}$), and absorption fraction ($f_{\text{abs}}$).  While magnetic field strength and wind speed both directly affect the total cosmic ray lifetimes, the ionized gas density and the absorption fraction only affect the emitted radio spectrum.  The free--free emission and absorption coefficients are both directly proportional to the square of $n_{\text{ion}}$, and so, the frequency at which the radio spectrum flattens or turns down and the amount of free--free emission at high frequencies both increase with $n_{\text{ion}}$.

\begin{figure*}
 \subfigure[East Nucleus, $\eta = 0.05$]{
  \includegraphics[width=0.49\linewidth]{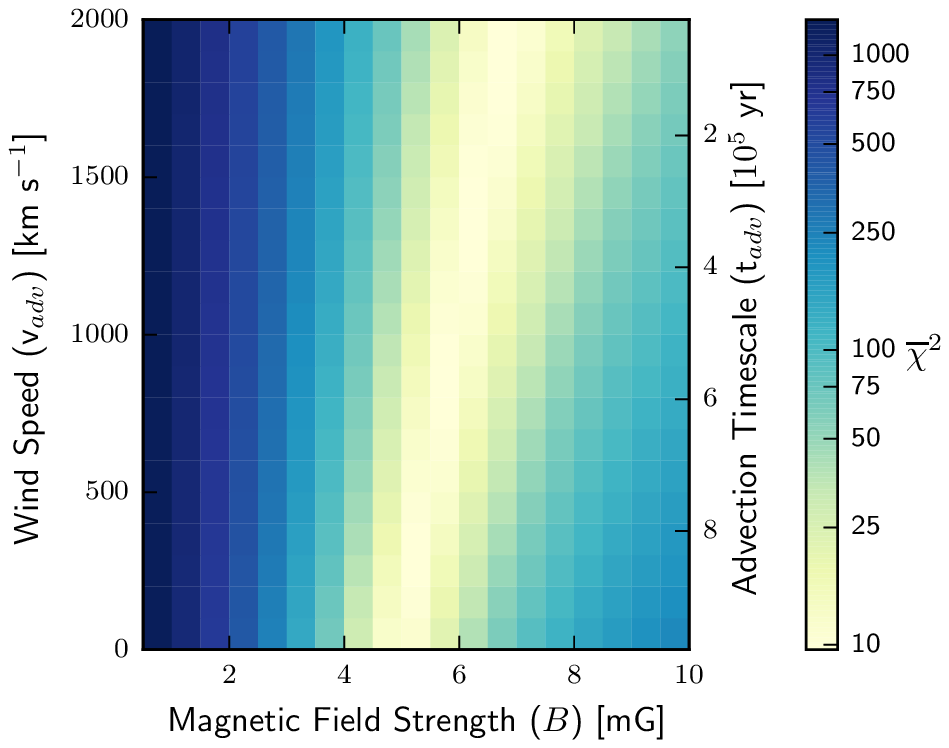}}
 \subfigure[West Nucleus, $\eta = 0.05$]{
  \includegraphics[width=0.49\linewidth]{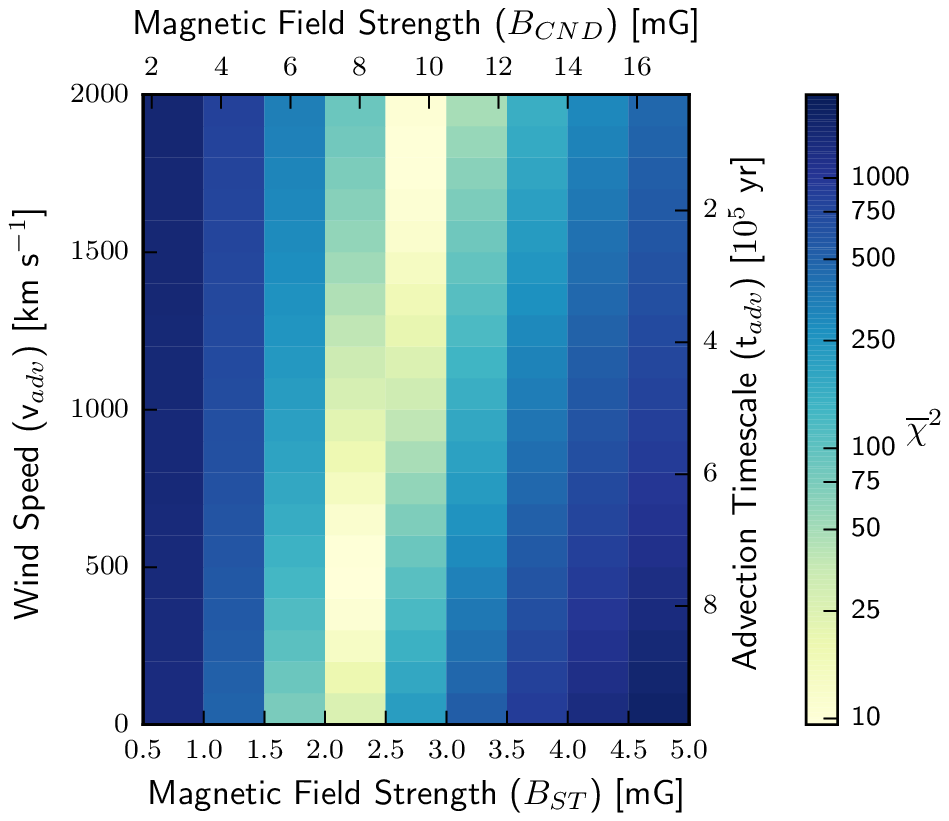}}
 \subfigure[East Nucleus, $\eta = 0.1$]{
  \includegraphics[width=0.49\linewidth]{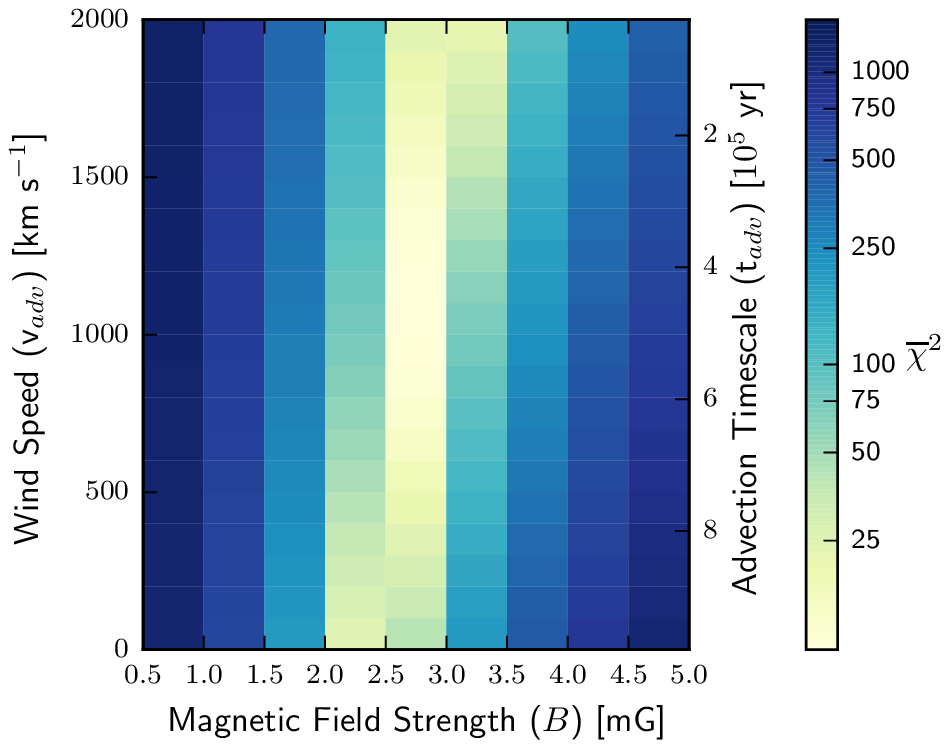}}
 \subfigure[West Nucleus, $\eta = 0.1$]{
  \includegraphics[width=0.49\linewidth]{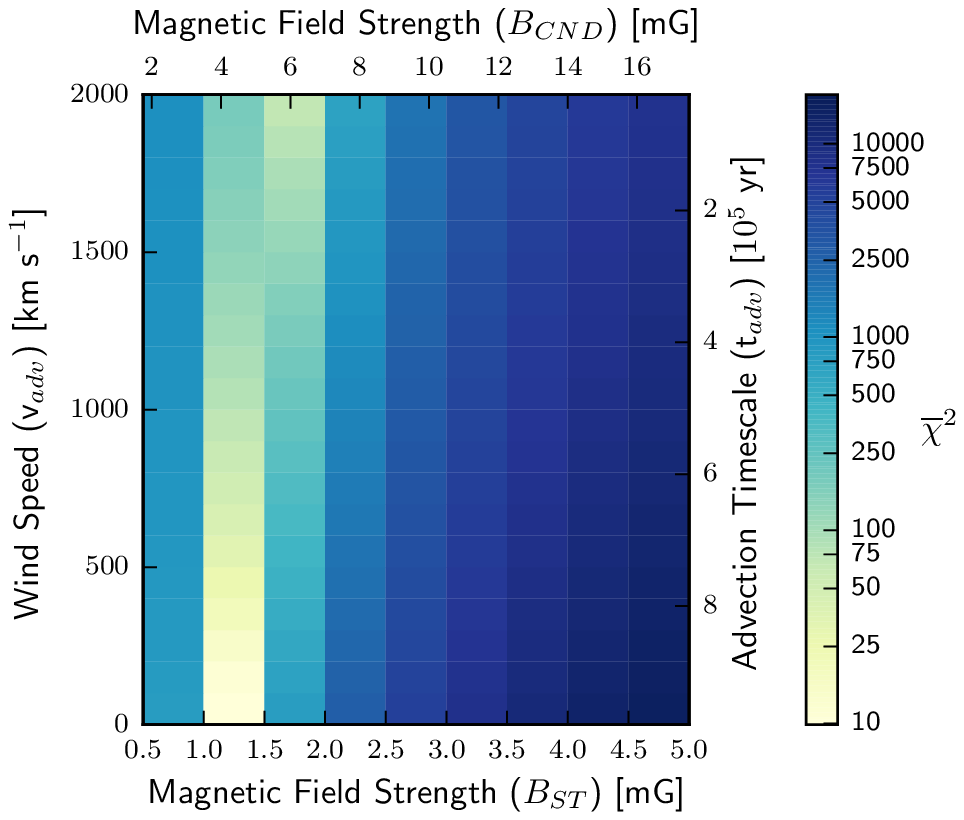}}
\caption{Contours of $\chi^{2}$ demonstrate that changes in wind speed have little to no impact on the radio spectrum.  Acceptable fits for no-escape (zero wind speed) cases indicate that the nuclei are proton calorimeters, although calorimetry fraction decreases with increasing wind speed.  The best-fitting models for the radio spectrum require milligauss strength magnetic fields and are limited to a single value for acceleration efficiencies of 10\%: 2.5~mG for the eastern nucleus and 1.0~mG for the western nucleus ST.  In the eastern nucleus, for the lower acceleration efficiency, a larger range of magnetic field strengths (4.5 -- 7.0 mG) produce models within $3\sigma$ of the best-fitting model and a degeneracy between magnetic field strength and wind speed is seen.}
\end{figure*}
\begin{figure*}
 \subfigure[East Nucleus]{
  \includegraphics[width=0.49\linewidth]{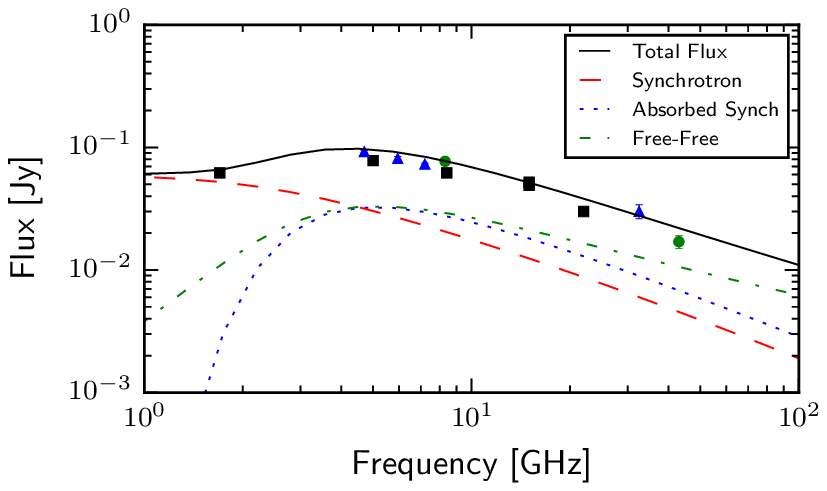}}
 \subfigure[West Nucleus]{
  \includegraphics[width=0.49\linewidth]{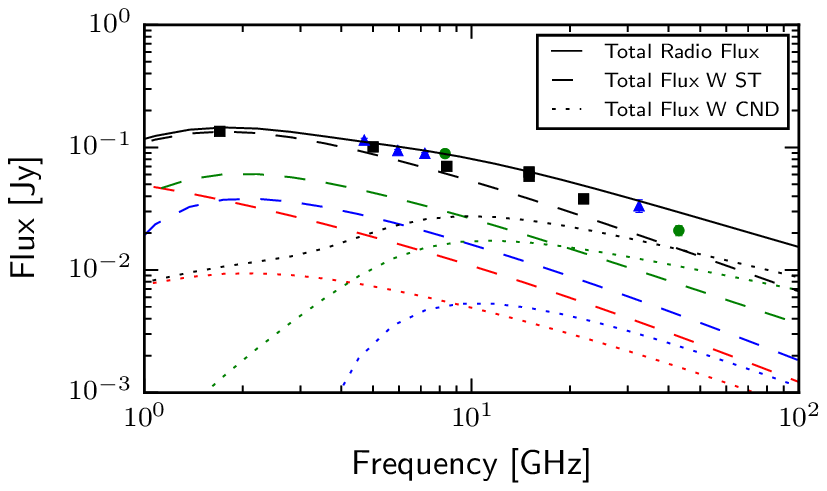}}
\caption{Best-fitting YEGZ models for the radio spectra in the eastern and western nuclei.  The total radio flux (black, solid line) in eastern nucleus includes non-thermal-synchrotron emission, both absorbed (blue, dotted line) and unabsorbed (red, dashed line), and free--free emission (green, dot--dashed line).  In the western nucleus, the total radio spectrum (black, solid line) incorporates emission from the inner CND (black, dotted line) and the ST (black, dashed line).  Data are taken from \citet[][black squares]{Downes98}, \citet[][green circles]{Rodriguez05}, and \citet[][blue triangles]{Barcos15}.}
\end{figure*}

Observations in \cite{Downes98}, \cite{Rodriguez05} and \cite{Barcos15} separate the integrated fluxes of the eastern and western nuclei from the total flux, allowing us to constrain parameters for each nucleus individually.  As we do not have radio observations which are separable between the two regions of the western nucleus, we cannot constrain the magnetic field strength in each region separately.  We therefore assume that the ratio between the magnetic field strength of the inner and outer regions of the western nucleus is equal to the square root of the ratio of the average gas densities, $B_{\text{CND}} / B_{\text{ST}} = \sqrt{n_{\text{CND}} / n_{\text{ST}}} \approx 3.5$ \citep{Crutcher12}.  Thus, our magnetic field strength determination for the innermost western nucleus is an estimate that is guided by Milky Way observations.

When assuming the standard 10\% cosmic ray acceleration efficiency, we find that agreement between the models and the observed radio data occurs only in a very narrow area of parameter space (see Fig. 2).  The best-fitting models for the nuclei have magnetic field strengths limited to 1.0 mG for the western nucleus (3.5 mG in the CND, estimated from scaling) and 2.0 -- 2.5 mG for the eastern nucleus (see Figs 2 and 3 and Table 2).  As seen in Fig. 3, the total radio emission in both nuclei flattens at low frequencies, and in the eastern nucleus, the radio spectrum may be turning over completely.  This flattening of the radio spectra requires moderate to high absorption fractions of 50 -- 100\% in the eastern nucleus and low to moderate absorption fractions of 10 -- 70\% in the western nucleus.

In addition to moderate absorption fractions in each nucleus, we also find a high contribution from thermal emission to the total radio spectrum (see Fig. 3), particularly in the western CND where the majority of the radio emission is thermal above $\sim$5 GHz.  In part, this unusually high fraction of thermal emission is due to the inability of the model to effectively fit for free--free absorption and free--free emission simultaneously as seen in the eastern nucleus and in previous work \citep[see YEGZ;][]{YoastHull14b}.  The ability of the models to accurately fit the fraction of thermal emission is further strained by the complicated nature of the western nucleus and the lack of separable radio observations.  Thus, in this particular case, the fractions of thermal emission in the best-fitting models have limited significance and do not necessarily contradict observations by \cite{Ana00} which indicate more modest amounts of thermal emission ($\sim 10-20\%$).

\subsection{Molecular Gas Mass}

The western nucleus Arp 220 has a very complex structure which we greatly simplified.  While our two-zone density distribution reproduces the observed peak column density of $\sim10^{25}$~cm$^{2}$, the mass is lower than that estimated by \cite{Scoville15} who derive the western nucleus gas mass from observations by the Atacama Large Millimeter Array (ALMA) of the submillimetre dust luminosity.  This yields a total gas mass of $4.2 \times 10^{9} ~ M_{\odot}$ for the western nucleus or four times our adopted value.  We therefore explored the effect of increased gas mass on our model by tripling the mass in the western torus and ran a limited suite models with fixed parameters.  We set the spectral index to $p = 2.3$ and ran $\chi^{2}$ tests over the entire range of magnetic field strengths and wind speeds but over a subset of the previously tested ionized gas densities and absorption fractions.  We ran tests on the western nucleus for acceleration efficiencies of 5\% and 10\%.

In comparing the results of these models with a larger gas mass, we find that none of the tested models are within $3\sigma$ of the best-fitting model at the lower assumed gas mass.  Further more, these results yield extremely short cosmic ray electron lifetimes such that the physical validity of the models are in question.  In addition to the higher gas mass estimates, \cite{Scoville15} also propose a geometry where the molecular gas in both nuclei is confined to a thin ($\approx$10~pc) disc.  The evolution of supernovae and cosmic ray interactions in this type of high molecular mass structure is beyond the scope of this study which is designed to estimate cosmic ray interaction rates in Arp~220 but will need to be considered when Arp~220 is detected in $\gamma$-rays.

\begin{figure*}
 \subfigure[$\gamma$-Ray Flux, East Nucleus]{
  \includegraphics[width=0.49\linewidth]{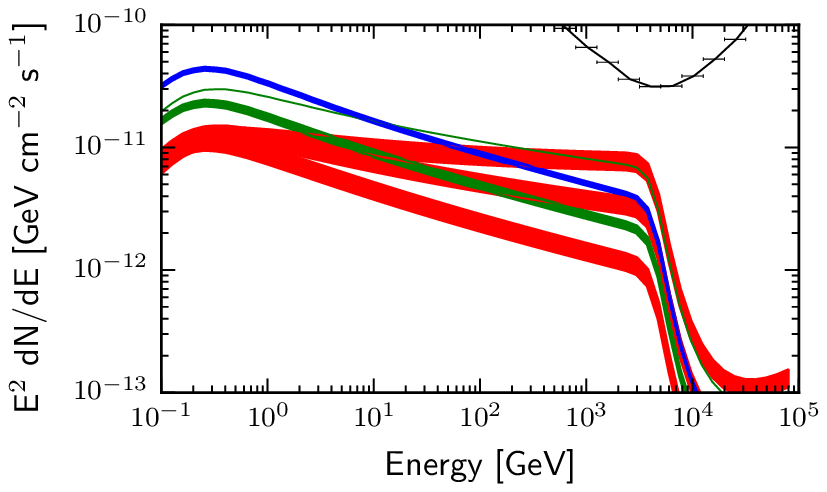}}
 \subfigure[$\gamma$-Ray Flux, West Nucleus]{
  \includegraphics[width=0.49\linewidth]{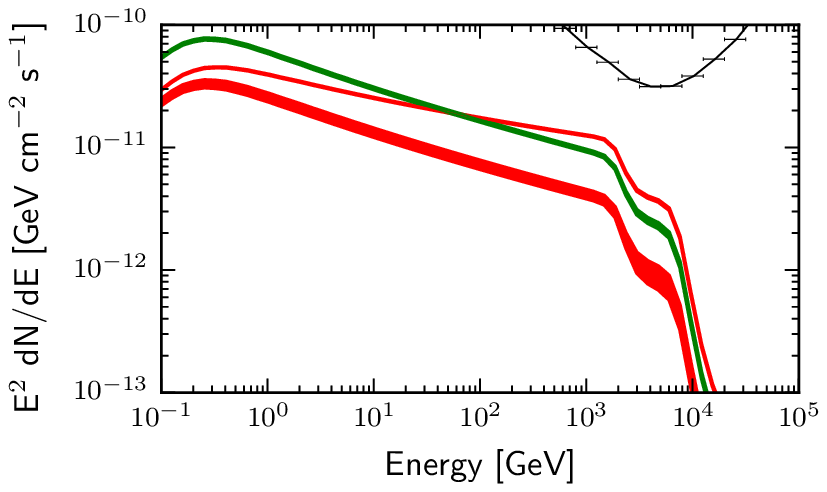}}
 \subfigure[$\gamma$-Ray Flux, Total]{
  \includegraphics[width=0.49\linewidth]{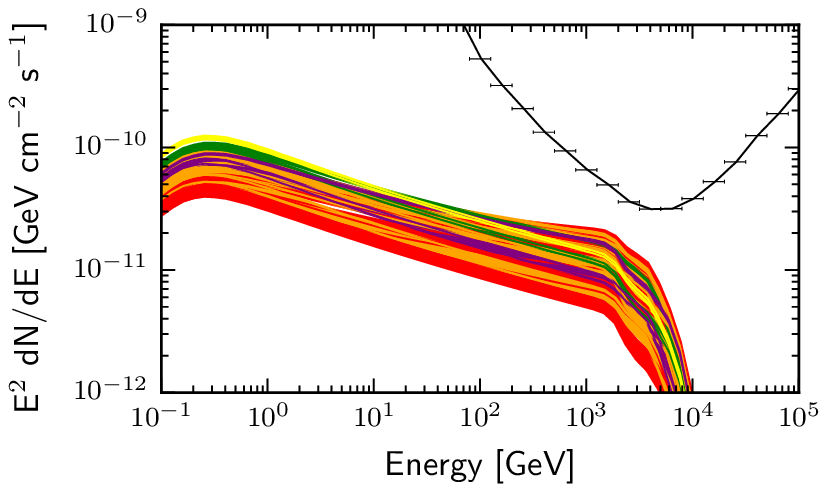}}
 \subfigure[Neutrino Flux, Total]{
  \includegraphics[width=0.49\linewidth]{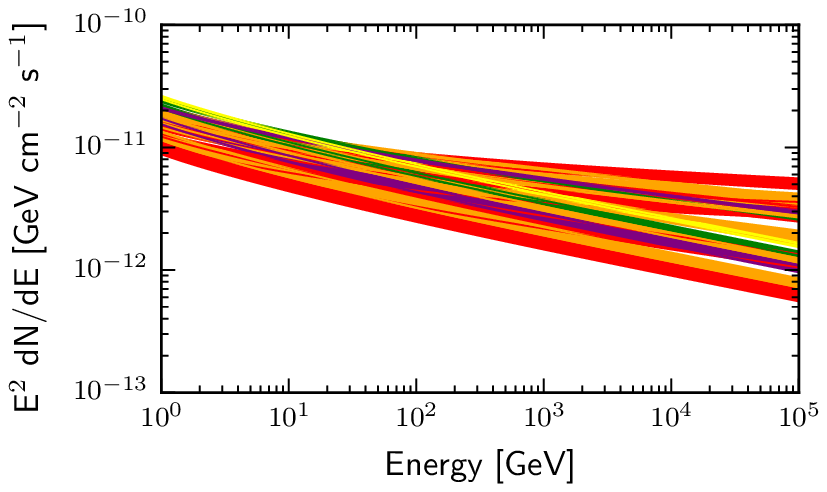}}
\caption{$\gamma$-ray and neutrino spectra for the parameters corresponding to those for models within $3\sigma$ of the best-fitting radio models.  The bands of spectra shown are displayed separately for each nucleus in the upper plots, where $\eta = 5\%$ is shown in red, $\eta = 10\%$ in green, and $\eta = 20\%$ in blue in the upper panel.  In the lower plots, each summation spectra the two nuclei are displayed such that $\eta_{E} = 5\%$ combined with $\eta_{W} = 5\%$ is shown in red, $\eta_{E} = 5\%$ combined with $\eta_{W} = 10\%$ and $\eta_{E} = 10\%$ combined with $\eta_{W} = 5\%$ are shown in orange, $\eta_{E} = 10\%$ combined with $\eta_{W} = 10\%$ is shown in green, $\eta_{E} = 20\%$ combined with $\eta_{W} = 10\%$ is shown in yellow, and $\eta_{E} = 20\%$ combined with $\eta_{W} = 5\%$ is shown in purple.  The black line shown on the $\gamma$-ray plots is the differential sensitivity limit for 50 h of observation with the southern CTA array, taken from https://portal.cta-observatory.org/Pages/CTA-Performance.aspx.}
\end{figure*}
%
%

\subsection{Acceleration Efficiency and Proton Calorimetry}

In our earlier works, we demonstrated that for a given ISM, the YEGZ models are highly sensitive to the total flux of cosmic rays \citep[YEGZ;][]{YoastHull14b}.  This flux is primarily effected by the original energy input into cosmic rays and the advective time-scale, or escape fraction.  The energy input into cosmic rays is determined by the supernova rate and the assumed acceleration efficiency ($\eta$).  Within the uncertainty in the supernova rate, we vary acceleration efficiency from 5 to 20\%.

As shown above, for the standard 10\% efficiency, the resulting best-fitting models are highly constrained in magnetic field strength, and we find that this is also true for an acceleration efficiency of 20\% (see Table 2).  However, for a lower acceleration efficiency of 5\%, equivalent to a lower supernova rate, we find a much larger range of acceptable fits in the eastern nucleus with magnetic field strengths ranging from 4 to 7.5 mG and wind speeds spanning our entire tested range.  As such, the best-fitting models for Arp 220 are essentially independent of wind (advection) speed (see Figs 1 and 2).

In contrast, a galactic wind was a vital component in modelling the cosmic ray populations of the starburst galaxies M82 and NGC 253 such that an extremely limited range of wind speeds resulted in fits within $3\sigma$ of the best-fitting models.  The wind speed determines the advective timescale for a galaxy and the fraction of cosmic rays which escape.  Thus, wind speed is intrinsically tied to the proton calorimetry fraction for a galaxy which is closely related to the total radio and $\gamma$-ray emission from a galaxy.  Other models for Arp 220 have assumed fixed advection time-scales, thus ensuring proton calorimetry with the high gas densities in Arp 220 \citep{Torres04,Lacki13}.  While our models agree with others in finding that the starburst regions of M82 and NGC~253 are only $\sim$40 -- 60\% proton calorimeters, we find that Arp 220's nuclei are 65 -- 100\% (eastern) and 90 -- 100\% (western CND) proton calorimeters (see Fig. 1).  

%
\begin{table*}
\begin{minipage}{170mm}
\begin{center}
\caption{Energy Density Distribution in Galaxies}
\begin{tabular}{lcccccc}
\hline
 & Supernova & Average Gas & Cosmic Ray & Radiation Field & Magnetic Field & Magnetic Field\\
 & Power & Density & Energy Density & Energy Density & Energy Density & Strength \\
 & (erg~yr$^{-1}$) & (cm$^{-3}$) & (eV~cm$^{-3}$) & (eV~cm$^{-3}$) & (eV~cm$^{-3}$) & ($\mu$G) \\
\hline
Milky Way & $2 \times 10^{48}$ & 1 & 1.4 & 0.3 & 0.9 & 6\\
M82 & $7 \times 10^{48}$ & 260 & 470 & 490 & 2200 & 300\\
Arp 220 East & $7 \times 10^{49}$ & 7700 & 1100 & 40 000 & $1.0 \times 10^{6}$ & 6500\\
Arp 220 West CND & $1.3 \times 10^{50}$ & 42 000 & 2500 & 440 000 & $1.2 \times 10^{6}$ & 7000\\
\hline
\multicolumn{7}{l}{Notes: Values for the Milky Way are taken from Table 1.5 in \citet{Draine11}.  For M82 and Arp 220, the values for the}\\
\multicolumn{7}{l}{cosmic ray and magnetic field energy densities taken from our best-fitting models (see Section 3 above and YEGZ).}
\end{tabular}
\end{center}
\end{minipage}
\end{table*}
%

\subsection{Future Detection of $\gamma$-Rays and Neutrinos}

Despite the uncertainty in the calorimetry fraction and the total cosmic ray flux in the eastern nucleus, we can still use our best-fitting models to make a prediction on the emitted $\gamma$-ray and neutrino fluxes from Arp 220.  To calculate the possible $\gamma$-ray flux, we apply the parameters of models within $3\sigma$ from our best-fitting radio model.  Combining each possible set of models from the eastern and western nucleus, we find that the resulting $\gamma$-ray spectra peak around $\sim$0.3 GeV with a maximum flux of $\sim10^{10}$ GeV~cm$^{-2}$~s$^{-1}$ (see Fig. 4). 

While this is roughly an order of magnitude lower than previous upper limits for Arp 220 \citep{Lacki11} and \textit{Fermi}'s differential sensitivity for four years of observations, it is only a factor of $\sim$2-3 times smaller than the flux level of the recently detected NGC~2146 \citep{Tang14}.  We also compared our $\gamma$-ray flux with the differential sensitivity 50 h of observations with the future southern CTA array (see Fig. 4) and find it to be only a factor of a few larger than our maximum flux.  Thus, Arp 220 may still be detectable by \textit{Fermi} within the next several years and is a good target for CTA, especially for energies near 1~TeV.

In addition to making a prediction for the $\gamma$-ray spectrum, we can use our same results from the radio emission to predict the neutrino flux from Arp 220.  Proton--proton interactions are responsible for the creation of secondary pions, both neutral and charged.  While the neutral pions decay into $\gamma$-rays, the charged pions decay into a neutrino and a muon which further decays into a secondary electron or positron and two more neutrinos.  The spectrum of the first neutrino from the decay of the charged pion is what we calculate here, as the calculation of the spectra of neutrinos produced during muon decay is more complex \citep[see][]{YoastHull14b}.  The flux of our maximum model is roughly $6 \times 10^{-12}$ GeV~cm$^{-2}$~s$^{-1}$ at 0.1 PeV, and at this energy, the range of possible models spans an order of magnitude in flux (see Fig. 4).  Current point source sensitivity limits for the northern sky for IceCube are $\sim10^{-9}$ GeV~cm$^{-2}$~s$^{-1}$, assuming a spectrum of $E^{-2}$ \citep{Aartsen14}.  Thus, it seems unlikely that Arp 220 will be detected as a point source during a similar time frame by IceCube.  However, extreme ULIRGs such as Arp 220 should make a significant contribution to a diffuse neutrino background \citep{Aartsen14b, Murase13}.

\subsection{$\gamma$--$\gamma$ Absorption}

In addition to accounting for $\gamma$-ray and neutrino emission in Arp 220, we have also take into account the effects of $\gamma$--$\gamma$ absorption due to the intense radiation fields in the nuclei \citep{Torres04}.  At TeV energies and above, $\gamma$-rays and infrared photons can interact to produce an electron / positron pair \citep{Dermer09, Bottcher12}.  The resulting electrons will be of TeV energies and most of their energy will be lost to emission of synchrotron X-rays \citep{Lacki13}.

Beginning at $\sim$2 -- 5 TeV, the opacity for $\gamma$--$\gamma$ absorption in both nuclei is significantly greater than 1.  This results in a steepening of the predicted $\gamma$-ray spectrum at high energies (see Fig. 4).  We find no such increase in slope in the neutrino flux as the steepening is an effect of interactions between the $\gamma$-ray and the ambient radiation field and not the cosmic ray proton population.  Therefore, in the case of Arp 220 and other such ULIRGs, the TeV $\gamma$-ray flux is an unreliable indicator of neutrino flux.  If the effects of spectral steepening by $\gamma$--$\gamma$ absorption are accurate, then Arp 220 is unlikely to be detected by CTA or other ground based Cherenkov telescopes above $\sim$10 TeV.  

\section{Discussion and Conclusions}

In applying the YEGZ models to Arp 220, we find that the central starburst regions of Arp 220 are moderate to complete cosmic ray proton calorimeters.  As such, the leptonic cosmic ray population is dominated by secondary electrons and positrons.  The majority of these secondaries are produced at low energies \citep[e.g.,][]{Lacki10b} and are likely a major contributor to heating of the ISM via ionization \citep{Papadopoulos10a,Papadopoulos11}.  

Based on our best-fitting models for the radio spectrum, we make predictions for both the $\gamma$-ray and neutrino fluxes.  Our maximum $\gamma$-ray spectrum is a factor of a 2 -- 5 lower than previous predictions by \cite{Torres04} and less than a factor of 2 lower than those by \cite{Lacki13}.  While the predicted $\gamma$-ray flux will likely be detected by \textit{Fermi} in the future, under our model assumptions Arp 220 is unlikely to be detected as a high energy neutrino point source with the current IceCube observatory.  Additionally, $\gamma$--$\gamma$ absorption of the TeV energy $\gamma$-rays make the TeV $\gamma$-ray flux a poor indicator of the neutrino flux in ULIRGs and other such systems with extremely intense infrared radiation fields.

In addition, we find that milligauss strength magnetic fields are still necessary to reproduce the observed radio fluxes from the starburst nuclei, even having assumed larger supernova rates than previous models by factors of 2 -- 5 \citep{Torres04, Lacki13}.  Differences in assumed volume and radiation field energy density across the various models account for the similar best-fitting magnetic field strengths despite the range in assumed supernova rates.  The origins of milligauss strength magnetic fields in extreme starbursts and their impact on the evolution of these systems merit further examination.

While the energy density in both magnetic and radiation fields is up from starbursts like M82 by two to three orders of magnitude, the change in the ratio of their energy densities is up by less than an order of magnitude (see Table 3).  Conversely, we see a much larger change in the ratio of magnetic field energy density to cosmic ray energy density.   Because the cosmic ray energy density depends on the particle energy loss rate, it does not increase at the same rate as the magnetic and radiation field energy densities (Yoast-Hull, Gallagher, Zweibel, in preparation).  Thus, the magnetic fields exceed energy equipartition with the cosmic rays by more than two orders of magnitude (see Table 3).

\section*{Acknowledgements}

This work was supported in part by NSF AST-0907837, NSF PHY-0821899 (to the Center for Magnetic Self-Organization in Laboratory and Astrophysical Plasmas), and NSF PHY-0969061 (to the IceCube Collaboration).  Part of this research was carried out during JSG's appointment as a Jubileumsprofessor at the Chalmers University of Technology.  We thank Susanne Aalto, Kazushi Sakamoto, Dave Sanders, Nick Scoville, and Eskil Varenius for conversations on Arp 220, Justin Vandenbroucke and Reinhard Schlickeiser for discussions regarding the modelling, and Francis Halzen for his help and support.  Additionally, we thank the referee for their helpful comments.

\appendix

\section{Two Zone Models}

Our single-zone model uses a simple solution to the radiative transfer equation of \citep[e.g.,][]{RL79,Bottcher12, Ghisellini13}
\begin{equation}
F_{\nu}^{\text{obs}} = F_{\nu}^{\text{int}} \times \frac{1 - e^{-\tau}}{\tau},
\end{equation}
where $F_{\nu}^{\text{int}}$ is the radiative flux prior to absorption, $F_{\nu}^{\text{obs}}$ is the radiative flux after absorption, and $\tau$ is the optical depth for either free-free absorption or $\gamma$-$\gamma$ absorption.  This is still the solution for the eastern nucleus and the surrounding torus in the western nucleus.  In the western CND, we must account for a standard emission and absorption region with an additional, external absorbing region.  This observed flux is given by
\begin{equation}
F_{\nu}^{\text{obs}} = \left( F_{\nu}^{\text{int}}  \times \frac{1 - e^{-\tau_{\text{inner}}}}{\tau_{\text{inner}}} \right) \times e^{-\tau_{\text{outer}}},
\end{equation}
where $F_{\nu}^{\text{int}}$ is still the radiative flux prior to absorption, $\tau_{\text{inner}}$ is the optical depth for $\gamma$-$\gamma$ or free-free absorption in the emission region, and $\tau_{\text{outer}}$ is the optical depth in the external, absorbing region.


%
\end{document}